\documentstyle[aps,prl]{revtex}
\def\stackunder#1#2{\mathrel{\mathop{#2}\limits_{#1}}}
\begin{document}
\title{Yang-Mills, Gravity, and String Symmetries}
\author{Thomas Branson$^1$ \\
Department of Mathematics\\
The University of Iowa\\
Iowa City, IA  52242 \\ 
\vskip.2in
R.P. Lano$^{2}$ \\
Centre for Theoretical Studies\\
Indian Institute of Science\\
Bangalore - 560 012\\
India\\
\vskip.2in
V.G.J. Rodgers$^{3}$\\
Department of Physics \\
and Astronomy\\
The University of Iowa\\
Iowa City, IA 52242-1479\\
}

\date{ To Appear in Physics Letters B}
\maketitle
\footnotetext[1]{branson@math.uiowa.edu}
\footnotetext[2]{ralph@hepmips.physics.uiowa.edu}
\footnotetext[3]{vincent-rodgers@uiowa.edu;  address correspondence}
\begin{abstract}
{\sf
In this work we  use constructs from the dual space of the 
semi-direct product of the Virasoro algebra and the affine Lie algebra 
of a circle to write a theory of gravitation which   
is a natural analogue of Yang-Mills theory.  The theory provides a 
relation between quadratic differentials in 1+1 dimensions 
and rank two symmetric tensors in higher dimensions as well as 
a covariant local Lagrangian for two dimensional  gravity.
The isotropy equations of coadjoint orbits are interpreted as
Gauss law constraints for a field theory 
in two dimensions, which enables us to
extend to higher dimensions.  The theory has a Newtonian limit in any 
space-time dimension. 
Our approach introduces a novel relationship between string theories 
and 2D field theories that might be useful in defining dual theories.
We briefly discuss how this gravitational field couples to fermions.} 
\end{abstract}


\section{Introduction}
Throughout the literature, the focus on coadjoint orbits 
\cite{kirrilov,loop,witten3} for two dimensional
theories has been in the construction of geometric actions that enjoy the
symmetries of an underlying Lie algebra. In fact, the celebrated 
WZW \cite{witten1} model
and Polyakov \cite{polyakov} gravity have been shown to be precisely
the geometric actions
associated with the affine Lie algebra of special unitary 
groups \cite{halpern,kac} and 
circle diffeomorphisms \cite{virasoro}
respectively \cite{rai,alek,weigman}. Recently we have turned our 
attention away from theories that live on the orbits and toward theories 
that are transverse to these orbits \cite{vgjr,lano2}.  This paper 
reports the latest progress in this direction. 
We promote the fixed coadjoint vectors to dynamical fields and  
reinterpret  the generator of isotropy on the orbits
as Gauss law constraints that arise from a 2D field theory.  
We should emphasize that this is not the method of coadjoint orbits that
is used to construct geometric actions.  In particular, the 2D gravitational 
action that we derive is not the Polyakov action \cite{polyakov}
since the Polyakov action is the anomalous contribution to the gravitational
action.  The analogy between our work and that in Ref.\cite{polyakov} 
is the same as the relationship between the 
Yang-Mills action and WZW action.  In our action, there is no a priori 
dimensional restriction.  The coadjoint vectors describe a  vector
potential associated with a gauge theory and another 
field (the gravitational tensor potential) which we interpret as a 
possible hitherto missing contribution to gravitation. 
The new carrier of gravitation is  a  
rank two symmetric tensor $T_{\mu \nu}\,$, that has units of mass squared. 
After gauge fixing in two dimensions, one finds residual coordinate 
transformations that explain the appearance of an 
inhomogeneous contribution.  
We are also able to extract the 
symplectic part of the gravitational Lagrangian.
The propagator then
has both a quadratic and quartic contributions  
 instead of  just quartic as believed earlier \cite{vgjr}. The final
action can live in {\em any dimension}, yet is consistent with the constraints 
that arise in the $2$ dimensional theory.  Since constraints 
in the 2 dimensional theory are directly related to symmetries of
a string theory, namely
the Virasoro algebra, our approach introduces a novel relationship between 
string theories and field theories that may be generalized for other 
algebras.  Furthermore this approach may provide insight into the dual 
relationship between string theories and two dimensional field theories.  
This theory of gravitation  might augment General Relativity 
with the inclusion of a 
gravitational tensor potential that can be used to study fluctuations in the 
gravitational field about a fixed  metric while maintaining
general covariance. However the details of this are unknown at present.
 Recovery of this putative new contribution
which reconciles the different characteristics of  two-dimensional 
and higher-dimensional gravitation
is the major thrust of this letter.  

{}To begin \cite{vgjr,lano2}, one treats the adjoint representation of the 
Virasoro algebra
and the affine Lie algebra with adjoint vector~${\cal F}=\left( \xi \left( 
\theta \right) ,\Lambda \left(
\theta \right) ,a\right)$ as
residual  coordinate 
and gauge transformations 
respectively on a gauge potential $A$ and a  gravitational potential $T$
which are components of a centrally extended coadjoint vector, 
{$ B=\left( { T}\left( \theta \right) ,{ A}\left(
\theta \right) ,\mu \right)$}.  The adjoint vector ${\cal F}$ then transforms 
$B$ through \cite{lano},
\begin{equation}
{ B}_{\cal F}=\left( \xi \left( \theta \right) ,\Lambda \left(
\theta \right) ,a\right) *\left( { T}\left( \theta \right) ,{ A}\left(
\theta \right) ,\mu \right) =\left( 
{ T}\left( \theta \right) _{{\rm new}%
},{ A}\left( \theta \right) _{{\rm new}},0\right) \mbox{,}
\label{coadjoint} 
\end{equation}
where the transformed  fields are defined by,
\begin{equation}
{ T}\left( \theta \right) _{{\rm new}}=
\;\stackrel{\rm new\;diff\;covector}{%
{}{}{\overbrace{\stackunder{coordinate\ trans}{\underbrace{2\xi ^{^{\prime
}}T+T^{^{\prime }}\xi +\frac{c\mu }{24\pi }
\xi^{\prime \prime \prime }}}-
\stackunder{gauge\ trans}{\underbrace{{\rm Tr}\left( A\Lambda^{\prime }\right) 
}}}}}  \label{codiff} 
\end{equation}
and
\begin{equation}
{ A(\theta )}_{\rm new}=\;\stackrel{\rm new\;gauge\;covector}
{\,\overbrace{%
\stackunder{ coord\ trans}{\underbrace{{ A}^{\prime }\xi
 +\xi^{\prime }{\rm A}}}-\stackunder{ gauge\ trans}{\,\underbrace{[\Lambda
\,{ A}-{\rm A\,}\Lambda ]+k\,\mu \,\Lambda^{\prime }}}}}.
\label{coaffine} 
\end{equation}

These transformations are to be thought of as gauge and coordinate 
transformations on a time-like slice of a $1+1$ dimensional field
theory.  They represent the residual transformations that are 
available after temporal gauge fixing.  
One recovers the {\em isotropy equations} of the coadjoint orbits defined by
the quadratic differential $T$, the one dimensional gauge field $A$, 
and the central extension $\mu$ by setting the quantities 
in  (\ref{codiff}) and
(\ref{coaffine}) to zero.  These equations will be used to  
extract the Gauss laws identifying the residual
symmetries after gauge fixing in 2D.  
The isotropy algebra will arise from field equations 
that have collapsed to Gauss law constraints when the action is
varied in two dimensions.
With this simple assertion we shall be able to write a theory 
for any dimension, and possibly augment Einstein's theory of 
General Relativity.

\section{Construction of the Action}
It is our intention  to educe a theory, analogous to 
Yang-Mills,
that corresponds to a gravitational action,
and which has a natural reduction 
to the quadratic differential $T$ when gauge fixed in 2D.   
Yang-Mills theory will
serve as an archetype in the construction of the gravitational contributions. 
{}To recapitulate our earlier assertion, we claim that the 
isotropy equations of the coadjoint orbits are just the residual gauge and 
coordinate transformations after fixing a temporal gauge in 2D.  These 
residual transformation are encoded in a Gauss law that arises from the 
field equations of one of the components of the gravitational tensor potential
in 2D. 

{}To begin the  construction of our action, we need to write a
Gauss generator that yields the time independent transformation 
laws that come from the isotropy equations of the orbits.  
Recall equations
(\ref{coadjoint}), (\ref{codiff}) and 
(\ref{coaffine}). 
\begin{equation}
\delta _{\cal F}\widetilde{B}=\{ \delta T,\delta
A,\delta \mu \} 
 = \{ 2\xi ^{^{\prime }}T+T^{^{\prime }}\xi +%
\frac{c\mu }{24\pi }\xi^{^{\prime \prime \prime }}-{\rm Tr}( A\Lambda
^{^{\prime }}) ,
A^{^{\prime }}\xi +\xi ^{^{\prime }}A+\left[ \Lambda A-A\Lambda
\right] +k\mu \,\Lambda ^{^{\prime }},\,0\}. \label{transforms} 
\end{equation}
Throughout we shall assume that the constant $k \mu = -1$. 
{}From here we can extract the transformation laws and
Gauss law constraints in 2D that lead directly to higher dimensional
theories which include interactions. 

Since we are interested in a picture of pure gravity we will consider 
just the coordinate transformations on $T$ and set the gauge field to 
zero.     
We claim that $T$ is the one remaining component of a rank two symmetric 
tensor
after gauge fixing in two dimensions.
This would be consistent with the fact that in $D$ dimensions there 
are $D$, $\xi^\alpha$ fields that serve as ``gauge parameters'' for 
coordinate transformations.  Therefore we can remove $D$ degrees of freedom.
    
This would be tantamount to setting $T_{\mu 0} = 0$, leaving
only the purely spatial components $T_{i j}$ as dynamical variables. 
Furthermore, for the two dimensional case, 
gauge fixing in the temporal gauge leaves an anomalous inhomogeneous 
transformation on $T_{1 1}$ or, using world-sheet notation, 
$T_{\theta \theta}$ that must be exhibited in a 
Gauss law constraint.

Since the 2D theory will be our bridge to higher dimensions,
let us focus on its structure.  
{}From (\ref{transforms}) we have $$\delta T= 2\xi ^{^{\prime
}}T+T^{^{\prime }}\xi + q~\xi ^{^{\prime \prime \prime }},$$
where $q={c\mu }/{24\pi }$.  
In this work
we stress that {\em $T_{\mu \nu}$ is a rank two tensor} and not a 
pseudo-tensor as we believed earlier \cite{vgjr}.   {\em The 
inhomogeneous
transformation of $T_{\mu \nu}$ is a consequence of gauge fixing in two 
dimensions}.
In 2D the field 
equations of the $T_{0 1}$ component  become constraints; 
as opposed to dynamical field equations.  

Our claim is that the above equation carries these residual 
gauge transformations after gauge fixing a 2D field theory.  
We need a Gauss law that will deliver the time independent 
coordinate transformations for the $T_{1 1}$
component.  
Let $X^{i j}$ be the conjugate variable to $T_{i j}$.   
This is a symmetric object whose transformation law will be determined 
by the Gauss law. 
Let
\begin{equation}
{(G_{\mbox {diff}})}_a = X^{l m} \partial_a T_{l m} 
-\partial_l(X^{l m} T_{a m}) -\partial_m (T_{l a} X^{l m}) 
- q \partial_a \partial_l \partial_m X^{l m}. \label{tict}
\end{equation}
This is the generator  of time independent coordinate transformations in $1+1$ 
dimensions with an additional inhomogeneous term. We
 have purposely  left in the tensor structure for future use in 
higher dimensions.
{}From the Poisson brackets, one can recover the transformations 
laws of $X^{a b}$ and $T_{a b}$.  We have
\begin{equation}
Q_{\mbox{diff}}=\int d^3x\ G_a \xi^a,  
\end{equation}
where the  $\xi^a$ are the time independent spatial 
translations.  As a result,
\begin{eqnarray}
\{Q_{\mbox{diff}},T_{l m} \}&=& -\xi^a \partial_a 
T_{l m} - T_{a m} \partial_l \xi^a-T_{l a} \partial_m \xi^a
 - q~\partial_a \partial_l \partial_m
\,\xi ^a = -\delta_\xi T_{l m} \nonumber \\
\{Q_{\mbox{diff}},X^{l m}\} &=&\xi^a \partial_a X^{ l m} - 
(\partial_a \xi^l) X^{a m} - (\partial_a \xi^m ) X^{l a} + 
(\partial_a \xi^a) X^{l m} = \delta_\xi X^{l m}.  \label{gaussdiff}
\end{eqnarray}
The gauge fixing conditions,  $T_{\mu 0} = 0,$ are preserved under the time
independent coordinate transformations.  

The transformation law for $X^{l m}$ defines it as a rank two 
tensor density of weight one.  Thus $X^{l m}$ can carry  a factor of 
$\sqrt{g}$ in its definition.  
In two dimensions, the transformation 
law for the ``space-space'' component of $X^{l m}$ reduces to the 
transformation law of elements in the
adjoint representation of the Virasoro algebra, viz.\ $X' \xi - \xi' X$.  
The fact that $X^{a b} $ is 
a tensor density follows since $Q$ must be a scalar.
  In what is to follow, we shall 
remove the $\sqrt{g}$ factor from the definition of $X$ and simply
multiply our Lagrangian density by the $\sqrt{g}$ factor.

We can now replace the fields $\xi^a$ with $T^{a 0}$ and write down the 
first part of the $1+1$ Lagrangian as: 

\begin{equation}
L_{0} = X^{l m}(T^{a 0} \partial_a T_{l m} + T_{ a m}\partial_l T^{a 0} 
+ T_{l a} \partial_m T^{a 0}). \label{10}
\end{equation}
Also, one can easily show that $L_{0}$ is invariant (up to total derivatives)
 under time independent 
spatial transformations in $1+1$ dimensions.    

As can be seen from the above gauge fixed expression, the conjugate 
momentum $X^{a b}$ comes from a tensor density  of the type 
$X^{\mu \nu}{}_{\rho}$, were
$\rho$ has been evaluated in the time direction, i.e $X^{a b} = X^{a b}{}_0$
in analogy with $E_i$ and $F_{i 0}$.  This tensor  is 
symmetric in its $\mu \nu$ indices.   
Furthermore, the appearance of spatial derivatives on $T^{a 0}$ in $L_0$
suggests that $X^{\mu \nu}{}_\rho$ comes from a covariant
tensor with the structure
$$
X_{\mu \nu \rho} =  \partial_\rho T_{\mu \nu}.
$$
On an arbitrary manifold with fixed metric we may write
the fully  covariant field $X_{\alpha \beta \gamma}$ as
\begin{equation}
X_{\alpha \beta \gamma} =  \nabla_\gamma T_{\alpha \beta}. 
\label{X} \nonumber \\
\end{equation}
Also we introduce the tensor
\begin{eqnarray}
K_{\gamma \alpha \beta} &=& \frac{1}{2}(\nabla_\gamma T_{\alpha \beta} - 
\nabla_\alpha T_{\gamma \beta}) \label{kappa}
\end{eqnarray}
which we will use later to couple $T_{\alpha \beta}$ to fermions.
The tensor $X_{\alpha \beta \gamma}$ is symmetric in its 
first two indices, while 
$K_{\alpha \beta \gamma}$ is also acyclic in all of its indices but 
anti-symmetric in its first two indices. 

For the moment let us set $q=0,$ eliminating the 2D inhomogeneous 
contribution to the constraint equation.
Then in any dimension  the 
following covariant version of Eq.(\ref{10}) can be used to recover the 
homogeneous part
of the constraint equation,  
\begin{equation}
{\cal L}_0 = X^{\lambda \mu \rho} Y_{\lambda \mu \rho},
\label{L0} 
\end{equation}
where
\begin{equation}
Y_{\lambda \mu \rho} = \big( T^\alpha{}_\rho \partial_\alpha T_{\lambda \mu}
+ T_{\lambda \alpha} \partial_\mu T^\alpha{}_\rho + T_{\alpha \mu} 
\partial_\lambda T^\alpha{}_\rho - \partial_\rho(T^\alpha{}_\lambda 
T_{\alpha \mu})\big). 
\end{equation}
Notice that the  tensor $Y_{\lambda \mu \rho}$ is covariant and
independent of the derivative operator or the metric 
used to raise and lower the indices of $T_{\alpha \beta}$.
This tensor represents a covariant extension  
of the Lie derivative used in Eq.(\ref{gaussdiff}) for
the derivation of $T_{\alpha \beta}$ on itself when $q=0$. 
Notice that if $T_{\alpha \beta}$ were a metric tensor, 
$Y_{\lambda \mu \rho}$
would vanish identically.   Therefore $Y_{\lambda \mu \rho}$ is unrelated to 
a connection and cannot be accessed by curvature.  
In a similar manner the inhomogeneous contributions ($q\ne 0$) to the 
constraints
in Eq.(\ref{tict}) are accessed through the Lagrangian
\begin{equation}
{\cal L}_1 = -\frac{q}4 {~}\nabla_\beta T^\beta{~}_\alpha \nabla_\lambda 
\nabla_\mu X^{\lambda \mu \alpha}. \label{L1}  
\end{equation}

By varying the sum of the above Lagrangians with respect to the $T_{\nu 0}$ 
components  and fixing the temporal gauge on a flat manifold we find
\begin{equation}
X^{l m 0} \partial_\nu T^{l m} - \partial_m (X^{m l 0} T_{l \nu}) 
-\partial_l(X^{m l 0} T_{m \nu}) -
q{~}\partial_\nu \partial_l \partial_m X^{l m 0}= 0. 
\end{equation}
In the above the Latin indices refer to space components and the Greek
space-time components.  
We see that we have recovered the generator of time independent coordinate 
transformations, Eq.(\ref{tict}).  In $1+1$  dimensions this can
be identified with the isotropy equation, $\xi T' + 2 \xi' T + q{~} \xi''' = 
0$ where $\xi$ takes the role of $X^{1,1}$ as an adjoint element (recall that 
it transforms as a tensor density).
  
However Eqs.(\ref{L0},\ref{L1}) cannot build the full action as they  do not 
determine the
symplectic structure.  Another contribution to the Lagrangian
should exists that confirms the relation between 
time derivatives of the dynamical variables and their conjugate momenta.
Consider the action,
\begin{equation}
S_{I} = \int \sqrt{g} \mu^{6-n} \left[ \left( -\nabla_\alpha 
T_{\beta \gamma} \right) X^{\beta \gamma \alpha} + {1\over 2}X^{\beta \gamma 
\alpha}
X_{\beta \gamma \alpha} \right] d^nx. \label{si}
\end{equation} 
Here $\mu$ is a gravitational coupling constant with inverse mass 
dimensions.  
$S_{I}$ is known only up to a scale factor, relative to $S_{0},$ since the 
field equations for $T_{\theta 0}$ have no contribution from $S_{I}$ in the
temporal gauge in two dimensions.  
It is obvious 
that $S_{I}$ contains that part of the Lagrangian 
which determines the symplectic structure of the action.
Since a variation of $S_{I}$ with respect to $\partial_0 T_{\alpha \beta}$
yields $X^{\alpha \beta 0}$ as the conjugate momentum.   
After gauge fixing only the $T_{i j}$ components have non-trivial 
conjugate momentum.   Furthermore, after gauge fixing, Eqs.(\ref{L0},\ref{L1}) 
will not contribute to the conjugate
momentum. 

Putting this all together and writing partial derivatives on $T_{\alpha 
\beta}$ in terms of $X$, Eq.(\ref{X}), we may write the fully coordinate 
invariant action in $n$ dimensions as
\begin{eqnarray}
S_{\mbox{AT}} =  &&\int d^nx \sqrt{g}~\mu^{8-n}
\big( X^{\lambda \mu \rho}~T^\alpha{}_\rho 
~X_{ \mu \lambda \alpha} +2 X^{\lambda \mu \rho}~ T_{\lambda \alpha}
~X^\alpha{}_{\rho \mu}  
- X^{\lambda \mu \rho} X^\alpha{}_{\lambda \rho} T_{\alpha \mu}
- X^{\lambda \mu \rho} X_{\alpha \mu \rho} T^\alpha{}_\lambda \big)   
\nonumber \\
 -\frac{q}4 && \int d^nx \sqrt{g}{}\mu^{8-n}
\left(X^{\alpha \beta}{}_ \beta {} \nabla_\lambda \nabla_\mu{} X^{\lambda \mu 
}{}_\alpha
\right)  \nonumber \\
- \frac12 &&\int d^nx \sqrt{g} \mu^{6-n} \left( X^{\beta \gamma \alpha} 
X_{\beta \gamma \alpha} \right).  \label{sat}
\end{eqnarray}
This action describes a self-interacting rank two field,
$T_{\alpha \beta}$ that in the very least defines classical gravity 
in two dimensions.  

Since the metric $g_{\alpha \beta}$ is omnipresent, we interpret
the field $T_{\alpha \beta}$ as further gravitational contributions about a 
background metric.  
It may provide a natural bifurcation between local gravitational effects 
and cosmological effects of gravity due to the background metrics. It may be 
that  without introducing any arbitrary splittings  or asymptotics of
the metric tensor, one can study fluctuations in gravity (due to 
diffeomorphisms) by studying $S_{\mbox{AT}}.$ The origins of the field 
$T_{\alpha \beta}$ are  from the diffeomorphism algebra and necessarily must 
be related to
gravitation. In two dimensions this theory along with the extended Polyakov 
action, Eq.(\ref{2daction}) (which is already gauge fixed), completely defines 
classical two dimensional gravity and its anomalous contributions coming from 
the effective action.
This is in direct analogy with the way that Yang-Mills and the WZW models 
complete 2D gauge theories.  

The procedure that we have developed is quite general and  might be useful in 
defining dual theories for string theories in two dimensions.
For any given algebra of the circle or string, one can construct the 
coadjoint orbits for the algebra and extract the anomalous contributions.
The coadjoint vectors serve as dynamical fields while the adjoint vectors 
serve as the conjugate momenta.  The isotropy equation for the orbits is then 
reinterpreted as a constraint equation for conjugate variables providing a 
field theory that will interpolate between the different coadjoint orbits.

\section{Matter Couplings}

It is natural to ask how $T_{\alpha \beta}$  might couple to matter.
Coupling to point particles is straightforward as one can write 
$(P^{\alpha}P^{\beta}) T_{\alpha \beta}$ and one can see that 
a Newtonian limit will exists in any space-time dimension.  
In order to couple this to fermions let us recall the 
2D geometric action found in \cite{rai,lano2,lano}. 
There  the $1+1$ dimensional theory admits the action 
\begin{equation}
S = \stackunder{\mbox{fermion-gravity}}{\underbrace 
{\int d^2 x~T(\theta){ \partial_\tau s \over \partial_\theta s}}} + 
\int d^3 x \left(\partial_\tau T~ { \partial_\lambda s \over
 \partial_\theta s} - \partial_\lambda T~ { \partial_\tau s \over
 \partial_\theta s}\right) 
+ {c \mu  \over 48\pi } \int  \left[
{{\partial^2_{\theta} s}\over{(\partial_{\theta}s)^2}} \partial
_{\tau} \partial_{\theta} s -  {{(\partial^2_{\theta}s)^2
(\partial_{\tau} s)}\over{(\partial_{\theta} s)^3}} \right] d\theta d\tau.
\label{2daction}
\end{equation} 
The bosonized fermions form the dimensionless rank two field, 
$\partial_\alpha s /
 \partial_\beta s $
and the fermion-gravity interaction dictates an interaction term, 
\begin{equation}
S_{\Psi T} = \int \sqrt{g} {\bar \Psi}\gamma^\mu \left( \partial_\mu + 
\omega_{\mu} +\mu^{2} K^{\alpha \beta}{}_{\mu} 
[\gamma_\alpha, 
\gamma_\beta] \right) \Psi d^4 x,
\end{equation}
where $K_{~~~\mu}^{\alpha \beta}$ is defined in (\ref{kappa}) and 
$\omega_{\mu}$ is the spin connection. Thus $K_{~~~\mu}^{\alpha \beta}$
acts as a  contribution to the  spin connection.  
The gauge fixed action of Eq.(\ref{sat}) along with its constraints
and Eq.(\ref{2daction}) define the theory of chiral fermions interaction with 
gravitation in two dimensions or if one prefers an effective
action of gravity in two dimensions with its anomalous contributions.
  
\vskip18pt

\section{Acknowledgments}
This work was partially supported by NSF grant PHY-94-11002, 
the Carver Foundation and the Obermann Center for Advanced Studies.
T.B.\ and V.G.J.R.\ are indebted to the
Obermann Center for Advanced Studies for
support and hospitality.  V.G.J.R.\ would like to thank the Institute for 
Fundamental Theory and the High Energy Experimental Group
at the University of Florida and the Aspen Center for Physics
where some of this work was done.  This work is dedicated to the 
memory of Emma Meurice.  
\eject


\begin{thebibliography}{99}
\bibitem{kirrilov} A.A. Kirillov, Lect. Notes in Math. 970 (1982) 101,
Springer-Verlag (Berlin)
\bibitem{loop}  A. Pressley and G. Segal, {\em Loop Groups}, Oxford 
Univ. Press (Oxford, 1986)
\bibitem{witten3} E. Witten, {\sl Comm. Math. Phy.}~{\bf 114} (1988) 1
\bibitem{witten1} E. Witten {\sl Comm. Math. Phys.}~{\bf 92} (1984) 455
\bibitem{polyakov} A.M. Polyakov, {\em Mod. Phys. Lett.}~{\bf A2}
(1987) 893
\bibitem{halpern} K. Bardakci and M.B. Halpern, {\sl Phys. Rev.}~{\bf D3} 
(1971) 2493
\bibitem{kac} V.G. Kac, {\sl J. Funct. Anal. Appl.}~{\bf 8} (1974) 68
\bibitem{virasoro} M.A. Virasoro, {\sl Phy. Rev. Lett.}~{\bf 22} (1969) 37
\bibitem{rai} B. Rai and V.G.J. Rodgers, {\sl Nucl. Phys.}~{\bf B341} 
(1990) 119\\
Gustav W. Delius, Peter van Nieuwenhuizen and V.G.J. Rodgers, 
{\sl Int. J. Mod. Phys.}~{\bf A5} (1990) 3943
\bibitem{alek}  A. Yu. Alekseev and S.L. Shatashvili,
 {\sl Nucl. Phys.}~{\bf B323} (1989) 719
\bibitem{weigman}  P.B. Wiegmann, {\sl Nucl. Phys.}~{\bf B323} (1989) 311
\bibitem{vgjr} V.G.J. Rodgers, {\sl Phys. Lett.}~{\bf B336} (1994) 343
\bibitem{lano2} Ralph Lano and V.G.J. Rodgers, {\sl Nucl. Phys.}~{\bf B437} 
(1995) 45
\bibitem{lano} R.P. Lano and V.G.J. Rodgers, {\sl Mod. Phys. Lett.}~{\bf A7} 
(1992) 1725 \\
R.P. Lano, ``{\em Application of Coadjoint Orbits to the Loop Group and the 
Diffeomorphism Group of the Circle},'' \\
Master's Thesis, The University of Iowa, May 1994
\end{thebibliography}
\end{document}